\documentclass[11pt,tightenlines,
    onecolumn,
    preprintnumbers,
    amsmath,
    amssymb,
    prd,
    showpacs
]{revtex4}
\usepackage{color}
\usepackage{graphicx}%
\usepackage{amsmath}
\usepackage[normalem,normalbf]{ulem}
\usepackage{amssymb}
\usepackage{bm}
\usepackage{enumerate}

\newcommand{\nn}{\nonumber}

\begin{document}
\preprint{CAS-KITPC/ITP-015}
\title{Slowly Rotating Charged Gauss-Bonnet Black holes in AdS Spaces}
\author{Hyeong-Chan Kim$^1$}
\email{hckim@phya.yonsei.ac.kr}
\author{Rong-Gen Cai$ ^2$}
\email{cairg@itp.ac.cn}
\affiliation{$^1$ Department of Physics, Yonsei University, Seoul
120-749, Republic of Korea \\
$^{2}$ Kavli Institute for Theoretical Physics China (KITPC) at
the Chinese Academy of Sciences,
and\\
Institute of Theoretical Physics, Chinese Academy of
Sciences, P.O. Box 2735, Beijing 100080, China }

\date{\today}%
\bigskip
\begin{abstract}
\bigskip
Rotating charged Gauss-Bonnet black hole solutions in higher
dimensional ($d >4$), asymptotically anti-de Sitter spacetime are
obtained in the small angular momentum limit. The angular
momentum, magnetic dipole moment, and the gyromagnetic ratio of
the black holes are calculated and it turns out that the
Gauss-Bonnet term does not affect to the gyromagnetic ratio.

\end{abstract}
\maketitle

\section{Introduction}

Due to the AdS/CFT correspondence~\cite{Mald,Gubs,Witten1},
over the past years a lot of attention has been focused on black
holes in anti-de Sitter (AdS) space. It was convincingly argued
by Witten~\cite{Witten2} that thermodynamics of black holes
in AdS spaces (AdS black holes) can be identified with that of a
certain dual conformal field theory (CFT) in high temperature
limit.  With this correspondence, one can gain some insights into thermodynamic properties and phase structures of strong coupling CFTs by studying thermodynamics of AdS black holes.

Nowadays it is well-known that the AdS Schwarzschild black hole
is thermodynamically unstable when its horizon radius is small,
while it is stable for large radius; there is a phase transition, named Hawking-Page phase transition~\cite{Hawk}, between the large stable black hole and a thermal AdS space. This phase transition is explained by Witten~\cite{Witten2} as the
confinement/deconfinement transition of Yang-Mills theory in
the AdS/CFT correspondence. Thus it is of interest to consider rotating/charged generalization of black holes in AdS spaces.
In the AdS/CFT correspondence, the rotating black holes in AdS
space are dual to  certain CFTs in a rotating space~\cite{Haw},
while charged ones are dual to  CFTs with chemical
potential~\cite{R-charged}. Indeed, the most general higher
dimensional rotating black holes in AdS space have been recently
found~\cite{Haw,Gibbons}.

On the other hand, it is also of interest to consider corrected
AdS black holes due to higher derivative curvature terms in the
low energy effective action of string theories. In the AdS/CFT
correspondence, these higher derivative curvature terms correspond
to the correction terms of large $N$ expansion in the CFT side.

Among the gravity theories with higher derivative curvature terms,
the so-called Gauss-Bonnet gravity is of some special features.
For example, first, the resulting field equations
contain no higher derivative terms of the metric than second order and it has been proven to be free of ghosts when expanding about the flat space, evading any problems with unitarity; second, the Gauss-Bonnet term appears in the low energy effective action of heterotic string theory; and third, the most important is that in the Gauss-Bonnet gravity, the analytic expression of static, spherically symmetric black hole solution can be found~\cite{Deser,Whee,Cai}.
Indeed, the Gauss-Bonnet term gives rise to some interesting effect on the thermodynamics of black holes in AdS space~\cite{Myers,Nojiri}.

It is of great interest to find some rotating black hole solutions
in the Gauss-Bonnet gravity. However, it is a rather hard task
since the equations of motion of the theory are highly nonlinear
ones. In this work, we would like to report on some results for
slowly rotating black hole solutions in the Gauss-Bonnet gravity
theory, here the rotating parameter appears as a small quantity.
Such so-called slowly rotating black holes have also been
investigated in other gravity theories~\cite{Horne}-\cite{Ghosh}.
Here we would like to mention that some rotating black brane
solutions have been obtained in the Gauss-Bonnet gravity theory in
\cite{Dehg}, but those so-called rotating solutions are
essentially obtained by a Lorentz boost from corresponding static
ones; they are equivalent to static ones locally, although not
equivalent globally.

This paper is organized as follows. In the next section, we first present the slowly rotating Gauss-Bonnet black hole solutions in AdS space.
The black hole horizon could be a surface  with positive, zero, or negative constant scalar curvature.
Some related thermodynamic quantities are calculated there.
In Sec.~III, we generalize to the charged case, and obtain the slowly
 rotating charged Gauss-Bonnet black hole solutions in AdS space.
Sec.~IV is a simple summary.

\section{Slowly Rotating Gauss-Bonnet Black Holes in AdS Space}

The Einstein-Hilbert action with a Gauss-Bonnet term and a
negative cosmological constant, $\Lambda=-(d-1)(d-2)/2l^2$, in $d$
dimensions can be written down as~\cite{Deser,Cai,Wilt}
\begin{equation}
\label{3eq1}
S=\frac{1}{16\pi G}\int d^dx\sqrt{-g}\left(R +\frac{(d-1)(d-2)}{l^2}
  + \alpha (R_{\mu\nu\gamma\delta}R^{\mu\nu\gamma\delta} -4 R_{\mu\nu}
   R^{\mu\nu}+R^2)-4\pi G F_{\mu\nu}F^{\mu\nu}\right),
\end{equation}
where $\alpha$ is the Gauss-Bonnet coefficient with dimension
$(length)^2$ and is positive in the heterotic string theory. We therefore restrict ourselves to the case $\alpha \ge 0$. The Gauss-Bonnet term is a topological invariant in four dimensions. Therefore $d\ge 5$ is assumed in this paper.
Further, $F_{\mu\nu}$ is the Maxwell field strength. Varying the action yields the equations of gravitational field
\begin{eqnarray}
\label{3eq2}
R_{\mu\nu}-\frac{1}{2}g_{\mu\nu}R &= &\frac{(d-1)(d-2)}{2l^2}g_{\mu\nu}
  + \alpha \left (\frac{1}{2}g_{\mu\nu}(R_{\gamma\delta\lambda\sigma}
  R^{\gamma\delta\lambda\sigma}-4 R_{\gamma\delta}R^{\gamma\delta}
  +R^2) \right. \nonumber \\
 &&~~~- \left. 2 RR_{\mu\nu}+4 R_{\mu\gamma}R^{\gamma}_{\ \nu}
  +4 R_{\gamma\delta}R^{\gamma\  \delta}_{\ \mu\ \ \nu}
   -2R_{\mu\gamma\delta\lambda}R_{\nu}^{\ \gamma\delta\lambda} \right) \nn \\
&&\quad +8\pi G \left(F_{\mu\alpha}F_\nu^{~\alpha}-\frac{1}{4}g_{\mu\nu}
    F_{\alpha\beta}F^{\alpha\beta}\right)   .
\end{eqnarray}
We assume the metric being of the following
form
\begin{equation}
\label{eq:metric}
ds^2 = -f(r)dt^2 +\frac{1}{f(r)}dr^2 +r^2h_{ij}dx^idx^j
    +2 a r^2 g(r)h(\theta) dt d\phi,
\end{equation}
where $a$ is a constant. $h_{ij}dx^idx^j$ represents the line
element of a ($d-2$)-dimensional hypersurface with constant
curvature $(d-2)(d-3)k$ and volume $\Sigma_k$, where $k$ is a
constant.
Without loss of generality, one may take $k=1$, $0$, and
$-1$, respectively. When $k=1$, one has
$h_{ij}dx^idx^j=d\theta^2+\sin^2\theta d\phi^2 +\cos^2\theta
d\Omega_{d-4}^2$ and $h=\sin^2\theta$; 
when $k=0$, $h_{ij}dx^idx^j=d\theta^2+d\phi^2 + dx_{d-4}^2$ and $h=1$; and
when $k=-1$, $h_{ij}dx^idx^j=d\theta^2+\sinh^2\theta
d\phi^2 +\cosh^2\theta d\Omega_{d-4}^2$ and
$h=\sinh^2\theta$, where $dx^2_{d-4}$ is the line element of a
$(d-4)$-dimensional Ricci flat Euclidian surface, while
$d\Omega_{d-4}^2$ denotes the line element of a
$(d-4)$-dimensional unit sphere.
Thus, the horizon of the black hole (\ref{eq:metric}) is a positive, zero, or negative constant scalar curvature surface, as $k=1$, $0$, and $-1$, respectively~\cite{Cai}.

In this section, we first consider the case without charge, namely
$F_{\mu\nu}=0$. When $g(r)=0$,  the function $f(r)$  describing a
static black hole solution was found in \cite{Cai}
\begin{equation}\label{eq4}
f(r)=k +\frac{r^2}{2\tilde\alpha}\left ( 1 -
 \sqrt{1-\frac{4\tilde \alpha}{l^2}}\sqrt{1+\frac{ m}{r^{d-1}}} \right),
\end{equation}
where $\tilde\alpha= (d-3)(d-4)\alpha$ and the integration
constant $m$ has a relation to  the gravitational mass $M$ of the
solution
\begin{eqnarray} \label{eq:m}
M= \frac{(d-2)\Sigma_k (1-4\tilde \alpha/l^2) m}{64 \pi G\tilde
\alpha}
 .
\end{eqnarray}
In the limit of $\tilde\alpha\rightarrow 0$, we have
\begin{eqnarray} \label{f:KerrAdS}
f_{\rm AdS}(r)= k+\frac{r^2}{l^2}-\frac{16 \pi G M}{(d-2)\Sigma_k}\frac1{
r^{d-3}} ,
\end{eqnarray}
which gives the AdS-Schwarzschild black hole solution with
positive, zero, or negative constant scalar curvature horizon,
depending on $k$. On the other hand, in the large $r$ limit,
$f(r)$ becomes
\begin{equation}\label{eq:f:asym}
f(r)=k +\frac{r^2}{l_{\rm eff}^2} - \frac{16\pi G M_{\rm
eff}}{(d-2) \Sigma_k }\frac{1}{r^{d-3}}  +O(r^{5-2d}),
\end{equation}
from which we can read off  the effective cosmological constant
and effective mass
\begin{eqnarray} \label{eq:leff}
\frac{1}{l_{\rm eff}^2}=\left ( 1 -
 \sqrt{1-\frac{4\tilde \alpha}{l^2}}\right) \frac{1}{2\tilde\alpha},
 \quad \quad M_{\rm eff} = \frac{M}{\sqrt{1-4\tilde\alpha/l^2}} .
\end{eqnarray}

\begin{figure}[htb]
  \includegraphics[width=.6\linewidth]{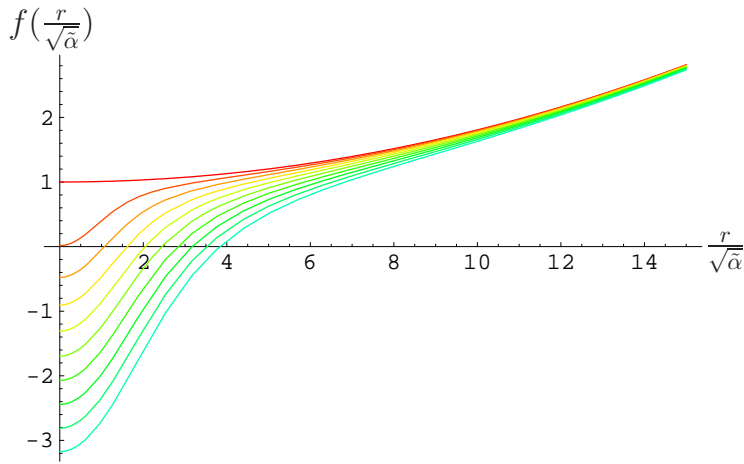}\\
  \caption{
  The Behavior of the function $f(x=r/\sqrt{\tilde\alpha})$ in the case of $k=1, ~d=5$, and $l^2=125\tilde\alpha$. The curves correspond to the mass parameter $m=\frac{7n+n^2}{2}\tilde\alpha^2$ with $n=0,1,\cdots ,10$, respectively from top to bottom.
   }\label{fig:fr}
\end{figure}
In Fig.~1 we plot the function $f(r)$ in the case of $d=5$ and
$k=1$. In this case, $f(r)$ is a pure increasing function of $r$
for $r> 0$. It approaches to the asymptotic form
$k+\frac{r^2}{2\tilde\alpha}(1-\sqrt{1-\frac{4\tilde\alpha}{l^2}})$
for large $r$. The metric has horizon if $m\geq 4 \tilde
\alpha^2$. Since the solution with $M=0$ is the AdS vacuum
solution, there is a mass gap from the AdS vacuum to the minimal
black hole with mass $m=4\tilde\alpha^2$. when $0<m<
4\tilde\alpha^2$, the solution describes a spacetime with a
deficit angle. When $d>5$, the mass gap disappears. When $k=0$ or
$k=-1$, the mass gap also disappears. For more details see
\cite{Cai}.

Now we consider the slowly rotating black hole solution with
$g(r)\ne 0$.  To the linear order of the parameter $a$, the metric
function $f(r)$ still keeps the form (\ref{eq4}). On the other
hand, the $t\phi$ component of equations of gravitational field leads to an
equation for the function $g(r)$
\begin{eqnarray} \label{eq:ddg}
\left(\log g'(r)\right)' = - \frac{2 d\, r^{d-1}+m (d+1)}{2 r \left(r^{d-1}+m\right)} .
\end{eqnarray}
It is interesting to note that this equation is independent of $k$.
After explicit integration, we have
\begin{eqnarray} \label{eq10}
g(r) =c_1+ \frac{2 c_2}{(d-1)m}
    \sqrt{1+\frac{m}{r^{d-1}}}\,,
\end{eqnarray}
where $c_1$ and $c_2$ are two integration constants. Now we 
fix these constants.  Expanding (\ref{eq10}) up to the
leading order of large $r$, one has
 \begin{equation}
  g(r)
=\left(c_1 + \frac{2c_2}{(d-1)m}\right)
    +\frac{c_2}{(d-1)r^{d-1}} +\cdots .
\end{equation}
Comparing this with the large $r$ asymptotic behavior of higher
dimensional Kerr-AdS solution given in \cite{Haw}, we find
\begin{equation}\label{eq:c12}
c_1= \frac{1}{2\tilde \alpha}, \quad c_2=
-\frac{(d-1)\sqrt{1-4\tilde \alpha /l^2}}{4\tilde \alpha}m .
\end{equation}
Then we obtain the function $g(r)$
\begin{eqnarray} \label{sol:gr:n}
g(r)=\frac{1}{2\tilde\alpha}\left ( 1 -
 \sqrt{1-\frac{4\tilde \alpha}{l^2}}\sqrt{1+\frac{ m}{r^{d-1}}} \right).
\end{eqnarray}
As a self-consistency check, we see that when $\tilde\alpha \to 0$, the solution (\ref{sol:gr:n}) indeed gives the asymptotic behavior of slowly rotating Kerr-AdS solution~\cite{Haw}.

For the slowly rotating solution, the horizon $r_+$ is still
determined by the equation $f(r)=0$, up to the linear order of the
rotating parameter $a$.  The coordinate angular velocity of a
locally nonrotating observer, with  four-velocity $u^\mu$ such
that $u\cdot \xi_{(\phi)}=0$, is
\begin{eqnarray} \label{eq:Omega}
\Omega=-\frac{g_{t\phi}}{g_{\phi\phi}} = a g(r)=
    \frac{a}{2\tilde\alpha}\left ( 1 -
 \sqrt{1-\frac{4\tilde \alpha}{l^2}}\sqrt{1+\frac{ m}{r^{d-1}}} \right).
\end{eqnarray}
In contrast to the ordinary Kerr black hole in asymptotically flat
spacetime, the angular velocity does not vanish at spatial
infinity in the present case.  Instead, we have the expression
\begin{eqnarray} \label{eq:Omega:infty}
\Omega_\infty = \frac{a}{2\tilde\alpha}\left(1-\sqrt{1-\frac{4\tilde\alpha}{l^2}}
    \right)=\frac{a}{l_{\rm eff}^2}.
\end{eqnarray}
We can see that only when $l\rightarrow \infty$, the angular
velocity vanishes at  spatial infinity. This is a remarkable
feature in AdS space. With $\tilde \alpha\rightarrow 0$, we also
get the correct Kerr-AdS limit.  On the other hand, the angular
velocity in~(\ref{eq:Omega}) on the horizon turns out to be
\begin{eqnarray} \label{eq:Omega:horizon}
\Omega_H= -\frac{k a}{r_+^2},
\end{eqnarray}
where we have used the fact that on the horizon $f(r_+)=0$. This
velocity $\Omega_H$ can be thought of as the angular velocity of
the black hole. One can also define the angular velocity of the
black hole with respect to a frame that is static at spatial
infinity. We have
\begin{eqnarray} \label{eq:omega}
\omega_H= \Omega_H-\Omega_\infty=-\left(\frac{a}{l^2_{\rm eff}}+
    \frac{a k}{r_+^2}\right).
\end{eqnarray}
Apparently, this angular velocity will vanish for $k=-1$ even for
nonzero $a$ when $r_+^2= 2\tilde \alpha$. However, this case will
not happen since in the case of $k=-1$, the minimal horizon of
black hole $r_{\rm min}^2>2\tilde\alpha$~\cite{Cai}. It is just
this angular velocity $w_H$ which enters into the first law of rotating black hole thermodynamics in AdS space~\cite{Gibb}.

The mass of the black holes can be expressed in terms of the
horizon radius $r_+$
\begin{eqnarray} \label{Mass}
M= \frac{(d-2)\Sigma_k r_+^{d-3}}{16\pi G} \left(k+
    \frac{\tilde \alpha k^2}{r_+^2}+\frac{r_+^2}{l^2}\right),
\end{eqnarray}
which is the same as the static one~\cite{Cai}. The  angular
momentum of the black hole is
\begin{eqnarray}
J =\frac{2 a M}{d-2} = \frac{a \Sigma_k r_+^{d-3}}{8\pi G}
    \left(k +\frac{\tilde \alpha k^2}{r_+^2}+\frac{r_+^2}{l^2}\right)
      .
\end{eqnarray}
The Hawking temperature of the black holes can be easily obtained
by requiring the absence of conical singularity at the horizon in
the Euclidean sector of the black hole solution.  It is the same
as the static case
\begin{eqnarray} \label{Temperatue}
T=\frac{(d-1) r_+^4+(d-3) k l^2 r_+^2 +(d-5) \tilde \alpha k^2 l^2}{
    4\pi l^2 r_+(r_+^2+ 2\tilde \alpha k)},
\end{eqnarray}
up to the linear order of the rotating parameter $a$,  since the
leading order of the parameter $a$ enters into the $g_{tt}$ metric component is second order, namely, $a^2$.  
From the first law of black hole thermodynamics $dM=TdS+\omega_H dJ$, one may easily see that the variation of the entropy is also second order in $a$. 
Therefore, to the linear order, the entropy expression will not be changed, the same as the static one~\cite{Cai}
\begin{equation}
S= \frac{\Sigma_k r_+^{d-2}}{4G}\left( 1+
\frac{d-2}{d-4}\frac{2\tilde \alpha k}{r_+^2}\right).
\end{equation}

\section{Slowly Rotating Charged Gauss-Bonnet Black Holes in AdS Space}

In this section, we consider the charged case. In this case, the
action, field equations, and the metric ansatz are the same as in Eqs.~(\ref{3eq1}), (\ref{3eq2}), and (\ref{eq:metric}), respectively. In the case of a charged static black hole, the spherical symmetry of the metric and the flux conservation gives us the electric field
\begin{eqnarray} \label{eq:Er:Ansatz}
-A_t'=F_{tr}=\frac{Q}{4\pi r^{d-2}} .
\end{eqnarray}
This gives the $A_t$ component of electro-magnetic potential
\begin{eqnarray} \label{A0}
A_t = \frac{Q}{4\pi(d-3) r^{d-3}} .
\end{eqnarray}
When $g(r)=0$, the function $f(r)$ describing a static charged black hole solution in $d-$dimensions is given by~\cite{Wilt,Nojiri}
\begin{eqnarray}\label{eq:f:r}
f(r) &=& k+\frac{r^2}{2\tilde\alpha}\left( 1 -
 \sqrt{1-\frac{4\tilde\alpha}{l^2}}\sqrt{1+\frac{m}{r^{d-1}}-\frac{q^2}{
    r^{2d-4}}} \right) ,\nn
\end{eqnarray}
where the gravitational mass $M$ and charge $Q$ of the solution
are
\begin{eqnarray} \label{eq:m}
M= \frac{(d-2)\Sigma_k (1-4\tilde \alpha/l^2)}{64 \pi\tilde \alpha }m, \quad\quad
Q^2=\frac{\pi(d-2)(d-3)(1-4\tilde \alpha/l^2)}{2\tilde \alpha G}q^2 .
\end{eqnarray}
\begin{figure}[htb]
  \includegraphics[width=.6\linewidth]{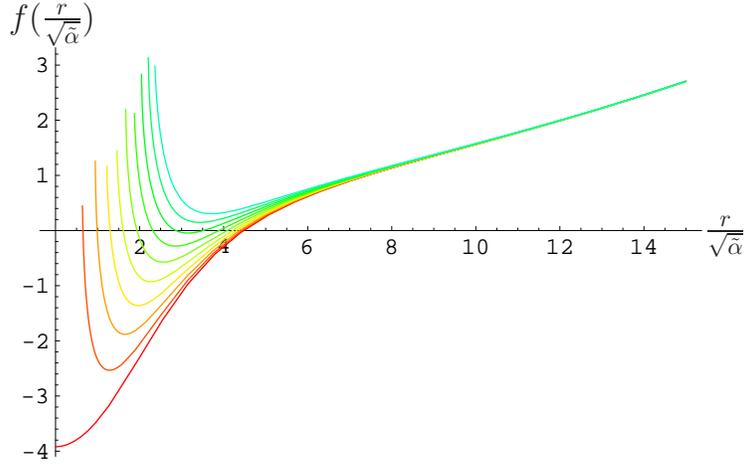}\\
  \caption{
  The behavior of the function $f(x=r/\sqrt{\tilde\alpha})$ for different $q$ in the case of $k=1, ~d=5$, $l^2=15\tilde\alpha$, and $m=100\tilde\alpha^2$.
   The curves corresponds to $q^2=5(7n+n^2)\tilde\alpha^3$ with $n=0,1,\cdots ,10$, respectively from the bottom to top.
   }\label{fig:frQ}
\end{figure}
The general behavior of the function $f(r)$ with respect to the variation of
 $Q$ is given in Fig.~\ref{fig:frQ}. The behavior of the solution is well
 analyzed in Ref.~\cite{Nojiri}.

Since the black hole rotates along the direction $\phi$, it
generates a magnetic field. To take into account this effect we
add the vector potential $A_\phi$
\begin{eqnarray} \label{eq:EMvector:Ansatz}
A_\phi = -a Q c(r) h(\theta) .
\end{eqnarray}
Then, the field equations of the electro-magnetic field
$\partial_a(\sqrt{-g} F^{ab})=0$, using the metric
ansatz~(\ref{eq:metric}), lead to an equation for function
$c(r)$
\begin{eqnarray} \label{eq:EMfield:eom}
(r^{d-4} f(r) c'(r))' -2k(d-3)r^{d-6} c(r)= \frac{g'(r)}{4\pi} .
\end{eqnarray}
Thus, once we obtain the metric $g(r)$, we can get the differential equation of electro-magnetic
 potential $c(r)$. In addition, up to the linear order of $a$,
 $A_t$ still keeps the form (\ref{A0}).

\begin{figure}[htb]
  \includegraphics[width=.6\linewidth]{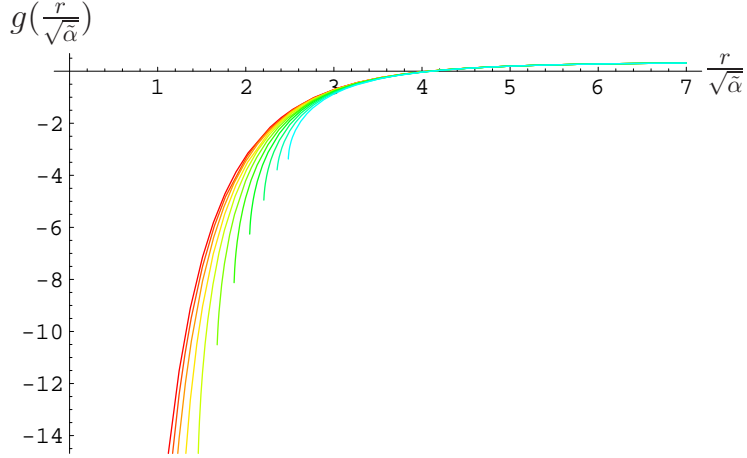}\\
  \caption{
 The behavior of $g(r)$ for  different $q$ with $d=5$, $l^2=15 \tilde \alpha$, $m=100\tilde\alpha^2$.
   $q$ is taken to be $5(7j+j^2)\tilde\alpha^3$
  with $j=0,\cdots 10$ from the bottom to top, respectively. $g(r)$ approaches to a constant $-\frac{1}{2\tilde\alpha}
  \left(1-\sqrt{1-\frac{4\tilde \alpha}{l^2}} \right) $ in the large $r$ limit.
   }\label{fig:gbar}
\end{figure}

As the case without charge, to the linear order of $a$, the metric
function $f(r)$ will not get correction from the rotation, that
is, the same as the static case. On the other hand, the $t\phi$
component of the field equations decouples from $c(r)$ and
leads to an equation for the function $g(r)$
\begin{eqnarray} \label{eq:ddg}
\left(\log g'(r)\right)'=-\frac{d}{dr}\left(\log r^d
\sqrt{1+\frac{m}{ r^{d-1}}-\frac{q^2}{r^{2d-4}}  }\right) .
\end{eqnarray}
Integrating this differential equation, we get a formal solution
\begin{eqnarray} \label{eq:gr}
g(r)&=&c_1-c_2\int\frac{ dr}{r^d \sqrt{1+\frac{m}{ r^{d-1}}-\frac{q^2}{r^{2d-4}}  }} \nonumber \\
   &=&c_1-\frac{c_2}{m}
    \int^{\frac{r}{m^{\frac1{d-1}}}} \frac{dx}{x^d\sqrt{1+\frac1{x^{d-1}}-
    \frac{q^2}{m^{\frac{2d-4}{d-1}}x^{2d-4}}}}   .
\end{eqnarray}
When $q=0$ we can explicitly integrate this integral to get the
result~(\ref{eq10}) given in the previous section
\begin{eqnarray}
g_{q=0}(r)=c_1+ \frac{2 c_2}{(d-1)m}
    \sqrt{1+\frac{m}{r^{d-1}}}\,.\nn
\end{eqnarray}
When $q \ne 0$, we are not able to give an explicit expression for $g(r)$. 
However, since the charge $q$ does not affect to the large $r$ behavior of the solution, the constants $c_1$ and $c_2$ can be determined, which are to be the same as Eq.~(\ref{eq:c12}), and therefore the solution $g(r)$ is
\begin{eqnarray} \label{gr:gen}
g(r) =\frac{1}{2\tilde\alpha}\left(1-
    \sqrt{1-\frac{4\tilde\alpha}{l^2}}\right)
   -\frac{(d-1)\sqrt{1-4\tilde \alpha/l^2}}{4\tilde \alpha}
    \int_{y}^\infty \frac{dx}{x^d\sqrt{1+\frac1{x^{d-1}}- \frac{q^2}{m^{\frac{2d-4}{d-1}}x^{2d-4}}}},
\end{eqnarray}
where $y=\frac{r}{m^{1/(d-1)}}$.
The asymptotic form of $g(r)$ is given by
\begin{eqnarray} \label{gr:asym}
g(r)
    &=& \frac{1}{l_{\rm eff}^2}-\frac{m\sqrt{1-\frac{4\tilde\alpha}{l^2}}}{
        4\tilde \alpha r^{d-1}} \left(1
   -\frac{m}{4 r^{d-1}} +\frac{(d-1)q^2}{2(3d-5)r^{2d-4}}+\cdots\right),
    \nn
\end{eqnarray}
where $g(r)$ approaches to $l_{\rm eff}^{-2}$.
We also can obtain small $q$ approximation of $g(r)$
\begin{eqnarray} \label{g:small q}
g(r) &=& \frac{1}{2\tilde\alpha}\left ( 1 -
 \sqrt{1-\frac{4\tilde \alpha}{l^2}}\sqrt{1+\frac{ m}{r^{d-1}}} \right) \\
 &-&\frac{m q^2\sqrt{1-4\tilde\alpha/l^2}}{4\tilde \alpha  \, r^{3d-5} }
 \frac{-1+\sqrt{1+ m/r^{d-1}} _2F_1(\frac{7d-11}{2(d-1)},
    \frac12, \frac{9d-11}{2(d-1)},-\frac m{r^{d-1}})}{
    \sqrt{1+r^{d-1}/m}}+O(q^4).\nn
\end{eqnarray}
On the horizon, up to ${\cal O}(q^2)$, it is
\begin{eqnarray} \label{grH}
g(r_+)&=& -\frac{k}{r_+^2}-\frac{q^2}{4\tilde\alpha}
    \frac{\sqrt{1-4\tilde\alpha/l^2}}{r_+^{2d-4} \sqrt{1+ \frac{m}{r_+^{d-1}}}}  \nn \\
  &-&\frac{m q^2\sqrt{1-4\tilde\alpha/l^2}}{4\tilde \alpha  \, r_+^{3d-5} }
 \frac{-1+\sqrt{1+ m/r_+^{d-1}} _2F_1(\frac{7d-11}{2(d-1)},
    \frac12, \frac{9d-11}{2(d-1)},-\frac m{r_+^{d-1}})}{
    \sqrt{1+r_+^{d-1}/m}}.
\end{eqnarray}
Therefore, the angular velocity of horizon $\Omega_H=a  g(r_+)$ also gets the corrections of ${\cal O}(q^2)$. As we see in
Fig.~\ref{fig:gbar}, $g(r_+)$ increases as $q$ increases. As a
result, the angular velocity of horizon also increases with $q$ for fixed $r_+$.

The function $c(r)$ is obtained by solving the differential
equation~(\ref{eq:EMfield:eom})
\begin{eqnarray} \label{diff:c}
(r^{d-4} f c')'-2k (d-3)r^{d-6} c= \frac{(d-1)m \sqrt{1-4\tilde\alpha/l^2}}{16\pi\tilde\alpha} \frac{1}{r^d\sqrt{1+\frac{m}{r^{d-1}}- \frac{q^2}{r^{2d-4}}}} .
\end{eqnarray}
We find that $c(r) $ can be written down as
\begin{eqnarray} \label{c:delta}
c(r)= -\frac{1}{4\pi (d-3) r^{d-3}} + q^2 \epsilon(r) .
\end{eqnarray}
Note that here the first term is independent of $k$. The function $\epsilon(r)$ satisfies
\begin{eqnarray} \label{eq:epsilon}
(r^{d-4} f \epsilon')'-2k(d-3)r^{d-6} \epsilon= \frac{(d-2) \sqrt{1-4\tilde\alpha/l^2}}{4\pi \,r^{2d-3}}
\frac{1}{\sqrt{1+\frac{m}{r^{d-1}}- \frac{q^2}{r^{2d-4}}}}.
\end{eqnarray}
We find that the leading term of $\epsilon$ is of the form
$\epsilon(r)\propto r^{-3d+7} $. Thus, in the large $r$ limit, we
arrive at
$$
c(r)\simeq -\frac{1}{4\pi (d-3) r^{d-3}} +\frac{ \sqrt{1-4\tilde\alpha/l^2}}{8\pi(3d-7)}\frac{q^2l_{\rm eff}^2}{ r^{3d-7}} .
$$
As a result, the electro-magnetic fields associated with the
solution are
\begin{eqnarray} \label{eq:Br}
F_{tr} = \frac{Q}{4\pi r^3},~~ F_{r\phi} = -a Q c'(r)h(\theta),~~
F_{\theta\phi} = -a Q c(r)h'(\theta).
\end{eqnarray}
 The expressions for the mass and the angular
momentum for this solution does not change through the
introduction of charge $Q$ since it does not alter asymptotic
behavior of the metric. The magnetic dipole moment for this slowly
rotating Gauss-Bonnet black hole is
\begin{eqnarray} \label{eq:MDM}
\mu= Q a.
\end{eqnarray}
Therefore, the gyromagnetic ratio is given by
\begin{eqnarray} \label{eq:gyro}
g= \frac{2\mu M}{Q J} = d-2
\end{eqnarray}
which depends only on the number of spacetime dimensions. The
value is the same as the case without the Gauss-Bonnet
term~\cite{Aliev}. In conclusion, the Gauss-Bonnet term does not
change the gyromagnetic ratio of the rotating black hole.

\section{Summary and Discussion}
Starting from the non-rotating charged Gauss-Bonnet black hole solutions in anti-de Sitter spacetime, we have obtained the slowly rotating solution by introducing a small angular momentum and solving the equations of motion up to the linear order of the angular momentum parameter.
If one chooses the metric $g_{t\phi}$ to be proportional to $r^2
g(r)$, the equation for $g(r)$ is much simplified as an integrable equation. 
The radial electric field is chosen so that the electric flux line to be continuous. The vector potential is chosen to have non-radial component $A_\phi=-aQ c(r)\sin^2\theta$ to represent
the magnetic field due to the rotation of the black hole. Since
the off diagonal component of the stress-tensor of electro-magnetic field is independent of $c(r)$, the equation for $g(r)$ decouples from $c(r)$ and is integrable.

As expected, our solution reduces to the slowly rotating Kerr-AdS black hole solution if the Gauss-Bonnet coefficient vanishes
$\tilde \alpha \rightarrow 0$. The expressions of the mass,
temperature, and entropy of the black hole solution, in terms of
the black hole horizon, do not change, up to the linear order of
the angular momentum parameter $a$. 
The angular momentum is written in terms of $a$ and the mass $M$ of the black hole and the gyromagnetic ratio of the Gauss-Bonnet black hole is obtained. 
It is shown that the Gauss-Bonnet term will not change the gyromagnetic ratio of the rotating black holes.

\begin{acknowledgments}
We thank K. Maeda and N. Ohta for helpful discussions.  This work
was supported in part by the Korea Research Foundation Grant
funded by Korea Government (KRF-2005-075-C00009; H.-C.K.). RGC was
supported in part by a grant from the Chinese Academy of Sciences,
and NSF of China under grants No.~10325525, No.~10525060 and
No.~90403029.

\end{acknowledgments} \vspace{1cm}


\begin{thebibliography}{10}
\bibitem{Mald} J.~Maldacena,
Adv.\ Theor.\ Math.\ Phys.\  {\bf 2}, 231 (1998) [Int.\ J.\
Theor.\ Phys.\  {\bf 38}, 1113 (1998)] [hep-th/9711200].

\bibitem{Gubs} S.~S.~Gubser, I.~R.~Klebanov and A.~M.~Polyakov,
Phys.\ Lett.\ B {\bf 428}, 105 (1998) [hep-th/9802109].


\bibitem{Witten1} E.~Witten,
Adv.\ Theor.\ Math.\ Phys.\  {\bf 2}, 253 (1998) [hep-th/9802150].


\bibitem{Witten2} E.~Witten,
Adv.\ Theor.\ Math.\ Phys.\  {\bf 2}, 505 (1998) [hep-th/9803131].

\bibitem{Hawk}S.~W.~Hawking and D.~N.~Page,
Commun.\ Math.\ Phys.\  {\bf 87}, 577 (1983).

\bibitem{Haw}S.~W.~Hawking, C.~J.~Hunter and M.~Taylor,
  Phys.\ Rev.\  D {\bf 59}, 064005 (1999)
  [arXiv:hep-th/9811056].

  \bibitem{R-charged}
A.~Chamblin, R.~Emparan, C.~V.~Johnson and R.~C.~Myers,
Phys.\ Rev.\ D {\bf 60}, 064018 (1999) [hep-th/9902170];
M.~Cvetic and S.~S.~Gubser,
JHEP {\bf 9904}, 024 (1999) [hep-th/9902195];
R.~G.~Cai and K.~S.~Soh,
  Mod.\ Phys.\ Lett.\  A {\bf 14}, 1895 (1999)
  [arXiv:hep-th/9812121];
S.~S.~Gubser,
  Nucl.\ Phys.\  B {\bf 551}, 667 (1999)
  [arXiv:hep-th/9810225].

\bibitem{Gibbons}G.~W.~Gibbons, H.~Lu, D.~N.~Page and C.~N.~Pope,
  Phys.\ Rev.\ Lett.\  {\bf 93}, 171102 (2004)
  [arXiv:hep-th/0409155];
G.~W.~Gibbons, H.~Lu, D.~N.~Page and C.~N.~Pope,
  J.\ Geom.\ Phys.\  {\bf 53}, 49 (2005)
  [arXiv:hep-th/0404008].


\bibitem{Deser}D.~G.~Boulware and S.~Deser,
Phys.\ Rev.\ Lett.\  {\bf 55}, 2656 (1985).

\bibitem{Whee}
J.~T.~Wheeler,
Nucl.\ Phys.\ B {\bf 268}, 737 (1986).

\bibitem{Cai}R.~G.~Cai,
  Phys.\ Rev.\  D {\bf 65}, 084014 (2002)
  [arXiv:hep-th/0109133];
R.~G.~Cai and Q.~Guo,
  Phys.\ Rev.\  D {\bf 69}, 104025 (2004)
  [arXiv:hep-th/0311020].

\bibitem{Myers}R.~C.~Myers and J.~Z.~Simon,
Phys.\ Rev.\ D {\bf 38}, 2434 (1988);
 R.~G.~Cai,
  Phys.\ Lett.\  B {\bf 582}, 237 (2004)
  [arXiv:hep-th/0311240];
 T.~Clunan, S.~F.~Ross and D.~J.~Smith,
  Class.\ Quant.\ Grav.\  {\bf 21}, 3447 (2004)
  [arXiv:gr-qc/0402044].
Y.~M.~Cho and I.~P.~Neupane,
  Phys.\ Rev.\  D {\bf 66}, 024044 (2002)
  [arXiv:hep-th/0202140];
I.~P.~Neupane,
  Phys.\ Rev.\  D {\bf 67}, 061501 (2003)
  [arXiv:hep-th/0212092];
  R.~G.~Cai and K.~S.~Soh,
  Phys.\ Rev.\  D {\bf 59}, 044013 (1999)
  [arXiv:gr-qc/9808067];
R.~G.~Cai,
  Phys.\ Rev.\  D {\bf 63}, 124018 (2001)
  [arXiv:hep-th/0102113];
M.~Banados, C.~Teitelboim and J.~Zanelli,
  Phys.\ Rev.\  D {\bf 49}, 975 (1994)
  [arXiv:gr-qc/9307033];
R.~Aros, R.~Troncoso and J.~Zanelli,
  Phys.\ Rev.\  D {\bf 63}, 084015 (2001)
  [arXiv:hep-th/0011097].

\bibitem{Nojiri}M.~Cvetic, S.~Nojiri and S.~D.~Odintsov,
  Nucl.\ Phys.\  B {\bf 628}, 295 (2002)
  [arXiv:hep-th/0112045].

\bibitem{Horne}J.~H.~Horne and G.~T.~Horowitz,
  Phys.\ Rev.\  D {\bf 46} (1992) 1340
  [arXiv:hep-th/9203083].
\bibitem{Shira}K.~Shiraishi,
  Phys.\ Lett.\  A {\bf 166}, 298 (1992).
\bibitem{GM1}T.~Ghosh and P.~Mitra,
  Class.\ Quant.\ Grav.\  {\bf 20}, 1403 (2003)
  [arXiv:gr-qc/0212057].
\bibitem{Shey}A.~Sheykhi and N.~Riazi,
  Int.\ J.\ Theor.\ Phys.\  {\bf 45}, 2453 (2006)
  [arXiv:hep-th/0605072].
\bibitem{Ghosh}
  T.~Ghosh and S.~SenGupta,
  arXiv:0709.2754 [hep-th].

\bibitem{Dehg} M.~H.~Dehghani,
  Phys.\ Rev.\  D {\bf 67}, 064017 (2003)
  [arXiv:hep-th/0211191];
 M.~H.~Dehghani,
  Phys.\ Rev.\  D {\bf 69}, 064024 (2004)
  [arXiv:hep-th/0312030];
M.~H.~Dehghani and R.~B.~Mann,
  Phys.\ Rev.\  D {\bf 73}, 104003 (2006)
  [arXiv:hep-th/0602243];
M.~H.~Dehghani, G.~H.~Bordbar and M.~Shamirzaie,
  Phys.\ Rev.\  D {\bf 74}, 064023 (2006)
  [arXiv:hep-th/0607067].

\bibitem{Wilt}D.~L.~Wiltshire,
Phys.\ Lett.\ B {\bf 169}, 36 (1986).

\bibitem{Gibb}G.~W.~Gibbons, M.~J.~Perry and C.~N.~Pope,
  Class.\ Quant.\ Grav.\  {\bf 22}, 1503 (2005)
  [arXiv:hep-th/0408217].

\bibitem{Aliev} A.~N.~Aliev,
  Phys.\ Rev.\  D {\bf 75}, 084041 (2007)
  [arXiv:hep-th/0702129].


\end{thebibliography}

\vspace{4cm}

\end{document}